\documentclass[10pt,conference]{IEEEtran}
\IEEEoverridecommandlockouts
\usepackage{cite}
\usepackage{url}
\usepackage{tikz}
\usepackage{amsmath,amssymb,amsfonts}
\usepackage{centernot}
\usepackage{amsthm}

\theoremstyle{definition}
\newtheorem{definition}{Definition}[section]
\newtheorem{example}{Example}[section]
\usepackage{algorithmic}
\usepackage[linesnumbered,ruled,vlined]{algorithm2e}
\usepackage{graphicx}
\usepackage{textcomp}
\usepackage{xcolor}
\usepackage{multirow}
\usepackage{bm}
\makeatletter
\usepackage[framemethod=TikZ]{mdframed}
\newcommand{\rmnum}[1]{\romannumeral #1}
\newcommand{\Rmnum}[1]{\expandafter\@slowromancap\romannumeral #1@}

\makeatother
\usepackage{newverbs}
\usepackage{fancyvrb}
\usepackage{adjustbox}
\usepackage{booktabs}
\usepackage{subfigure}
\usepackage{amssymb}
\usepackage{pifont}
\newcommand{\cmark}{\ding{51}}%
\newcommand{\xmark}{\ding{55}}%

\definecolor{liytpurple}{HTML}{008B8B}
\definecolor{b_blue}{HTML}{00008B}
 \definecolor{cRed}{HTML}{C00001}
\newcommand{\lyt}[1]{{{#1}}}

\newcommand{\chm}[1]{{{#1}}}
\newcommand{\xzw}[1]{{{#1}}}

\newcommand{\czx}[1]{{{#1}}}
\newcommand\correspondingauthor{\thanks{$^*$Corresponding author.}}
\usepackage{fancyvrb}
\usepackage{newverbs}
\usepackage{fancyvrb}
\usepackage{adjustbox}
\newverbcommand{\cverb}{\color{cRed}}{\underline{}}

\mdfdefinestyle{MyFrame}{%
    linecolor=black,
    outerlinewidth=0.3pt,
    roundcorner=5pt,
    innertopmargin=2.5pt, 
    innerbottommargin=2.5pt, 
    innerrightmargin=2.5pt,
    innerleftmargin=2.5pt,
    backgroundcolor=gray!25!white}

\def\BibTeX{{\rm B\kern-.05em{\sc i\kern-.025em b}\kern-.08em
    T\kern-.1667em\lower.7ex\hbox{E}\kern-.125emX}}
\begin{document}

\title{
{\scshape TransRegex}: Multi-modal Regular Expression Synthesis by Generate-and-Repair
}
\author{%
Yeting Li$^{\dag}$$^{\S}$, Shuaimin Li$^{\S}$, Zhiwu Xu$^{\natural}$, Jialun Cao$^{\ddag}$, Zixuan Chen$^{\dag}$$^{\S}$, Yun Hu$^{\sharp}$$^{\S}$, Haiming Chen$^{\dag*}$\correspondingauthor,  Shing-Chi Cheung$^{\ddag}$

\\
\IEEEauthorblockA{$^{\dag}$ State Key Laboratory of Computer Science, Institute of Software, Chinese Academy of Sciences, Beijing, China}
\IEEEauthorblockA{$^{\S}$ School of Computer Science and Technology, University of Chinese Academy of Sciences, Beijing, China}
\IEEEauthorblockA{$^{\natural}$ College of Computer Science and Software Engineering, Shenzhen University, Shenzhen, China}
\IEEEauthorblockA{$^{\ddag}$ The Hong Kong University of Science and Technology, Hong Kong, China}
\IEEEauthorblockA{$^{\sharp}$ Science \& Technology on Integrated Infomation System Laboratory,\\ Institute of Software, Chinese Academy of Sciences, Beijing, China}
\IEEEauthorblockA{$^{\dag}$\{liyt,chenzx,chm\}@ios.ac.cn,
$^{\S}$lishuaimin17@mails.ucas.ac.cn\\
$^{\natural}$xuzhiwu@szu.edu.cn,
$^{\ddag}$\{jcaoap,scc\}@cse.ust.hk\\ $^{\sharp}$huyun2016@iscas.ac.cn}
}


\maketitle

\begin{abstract}
\chm{Since regular expressions (abbrev. {\it regexes}) are difficult to understand and compose, automatically generating regexes has been an important research problem. This paper introduces {\scshape TransRegex}, for automatically constructing regexes from both natural language descriptions and examples. To the best of our knowledge, {\scshape TransRegex} is the first to treat the NLP-and-example-based regex synthesis problem as the problem of NLP-based synthesis with regex repair.
For this purpose, we present novel algorithms for both NLP-based synthesis and regex repair.
We evaluate {\scshape TransRegex} with ten \lyt{relevant} \iffalse related\fi  state-of-the-art tools on three publicly available datasets. The evaluation results demonstrate that the accuracy of our {\scshape TransRegex}  is 17.4\%, 35.8\% and 38.9\% higher than that of NLP-based approaches on the three datasets, respectively. Furthermore, {\scshape TransRegex} can achieve higher accuracy than the state-of-the-art multi-modal techniques with 10\% to 30\% higher accuracy on all three datasets. The evaluation results also indicate {\scshape TransRegex} utilizing natural language and examples in a more effective way.}
\end{abstract}

\begin{IEEEkeywords}
regex synthesis, regex repair, programming by natural languages, programming by example
\end{IEEEkeywords}

\section{Introduction}\label{sec1}

As \chm{a} versatile \chm{mechanism} for pattern matching and searching, regular expressions (abbrev. {\it regexes}) have been widely used in different fields of computer science such as programming languages, natural language processing (NLP) and databases due to the high effectiveness and accuracy\lyt{~\cite{ase2019,DBLP:conf/kbse/Shen000ML18,DBLP:conf/kbse/ChapmanWS17,DBLP:journals/tkde/BartoliLMT16,DBLP:conf/sigsoft/DavisCSL18,DBLP:conf/sigsoft/DavisMCSL19}}.
Unfortunately, despite their popularity, regexes can be difficult to understand and compose even for experienced
programmers~\cite{DBLP:conf/kbse/ChapmanWS17,DBLP:conf/ecoop/SpishakDE12,DBLP:conf/hase/LiuJW19,ase2019,DBLP:conf/acl/FengHZYLW18}.

\chm{To alleviate this problem, prior research has proposed techniques to automatically
generate regexes. For example, several techniques generate regexes from natural language  \lyt{(NL)} descriptions~\cite{DBLP:conf/naacl/KushmanB13,DBLP:conf/emnlp/LocascioNDKB16,DBLP:conf/emnlp/ZhongGYPXLLZ18,park-etal-2019-softregex}, while others \lyt{synthesize} regexes
from examples~\cite{DBLP:conf/oopsla/WangGS16,DBLP:conf/pldi/GulwaniJTV11,DBLP:journals/tods/BexNSV10,DBLP:journals/tweb/BexGNV10,DBLP:journals/mst/FreydenbergerK15,DBLP:conf/gpce/LeeSO16,ase20}. Though these
techniques help \lyt{lessen} \iffalse alleviate\fi the difficulties of automatic regex synthesis, they have obvious drawbacks as follows.} 

\chm{Existing NLP-based techniques can only generate regexes similar in shape to the training data and have relatively low accuracy on simple benchmark datasets (e.g., {\scshape SoftRegex}~\cite{park-etal-2019-softregex} achieved only 62.8\% accuracy on benchmark NL-RX-Turk~\cite{DBLP:conf/emnlp/LocascioNDKB16}). Furthermore, NLP-based techniques are impeded by the ambiguity and imprecision of  \lyt{NL} even for stylized English~\cite{pldi20,park-etal-2019-softregex,DBLP:conf/aaai/ZhongGY0LLZ18}. For example, according to \cite{park-etal-2019-softregex}, among 921 incorrectly predicted regexes, over 38.4\% are caused by ambiguity of  \lyt{NL} descriptions and 27.8\% are from imprecision.
Additionally, Zhong et al.~\cite{DBLP:conf/aaai/ZhongGY0LLZ18} found that NLP-based techniques may not make correct prediction if words in these  \lyt{NL} \iffalse natural language\fi descriptions are not covered by the training data.}

\chm{On the other hand, the example-based synthesis approaches rely on high quality examples provided by users. The synthesized regexes may be under-fitting or over-fitting when the given examples do not meet the implicit quality requirements (e.g, insufficient or not characteristic enough). However, examples with high quality are often \lyt{unavailable} in practice. It poses difficulties in applying \lyt{purely} example-based approaches. In addition, these approaches have severe restrictions on the kinds of regexes that they can synthesize (e.g., absence of Kleene star~\cite{DBLP:conf/oopsla/WangGS16,DBLP:conf/pldi/GulwaniJTV11}, limited character occurrences~\cite{DBLP:journals/tods/BexNSV10,DBLP:journals/tweb/BexGNV10,DBLP:journals/mst/FreydenbergerK15,ase20} or constraining to binary alphabet~\cite{DBLP:conf/gpce/LeeSO16}).}

\chm{Therefore,  to better synthesize regexes it would be ideal to take advantage of both  \lyt{NL} and examples (called NLP-and-example-based synthesis or multi-modal synthesis): the use of advanced NLP-based techniques can reduce the amount of  required \lyt{(characteristic)} examples  meanwhile \chm{alleviate} the amount of effort from users; while the use of examples can effectively disambiguate or correct errors in the descriptions. Further, a survey on posts on regex \lyt{synthesis} shows that many programmers actually use NL descriptions as a major resource, and leverage some example(s) to resolve the ambiguities of  \lyt{NL}~\cite{DBLP:conf/aaai/ManshadiGA13}.
On the other hand, \lyt{insufficient (characteristic) examples limit the generalization ability of example-based approaches, while incorporating NL can improve the generalization ability, and help to drastically narrow down the search space~\cite{DBLP:conf/aaai/ManshadiGA13}.}
Actually there have been recent attempts in this direction~\cite{pldi20,tacl}, in which they first translated the  \lyt{NL} description into a sketch\footnote{A sketch is an incomplete regex containing holes to denote missing components.}, then searched the regex space defined by the sketch guided by the given examples. However, the forms of translated sketches are restricted. This prevents regexes from being synthesized correctly when the generated sketches are inappropriate (e.g., logically-incorrect). In such a case, the incorrectness will be inherited from sketches to the subsequent regex.
Moreover, while these works~\cite{pldi20,tacl} have achieved relatively high accuracy on simple datasets, they did not perform well on complex and realistic datasets.
In latter datasets, the  \lyt{NL} \iffalse natural language\fi descriptions are longer, more complicated, and describe the regexes which are more complex in terms of length and tree-depth~\cite{DBLP:conf/acl/YeCDD20,DBLP:conf/aaai/ZhongGY0LLZ18}.}

We observe that most of the incorrect regexes generated by NLP-based techniques are very similar to the target regexes with subtle differences, and can be made equivalent\footnote{Two regexes are equal iff their corresponding languages are equivalent.} to
the target regexes with only minor modifications (e.g., reordering/revising characters or quantifiers).
\chm{This motivates us to view the NLP-and-example-based regex synthesis problem \iffalse the problem of synthesizing regexes from both  \lyt{NL} \iffalse natural language\fi and examples\fi as the problem of NLP-based synthesis with regex repair, and}
develop \chm{the first} framework, {\scshape TransRegex}, to leverage both  \lyt{NL} and examples
for regex synthesis \chm{by using NLP-based and regex repair techniques}.
{\scshape TransRegex} uses an NLP-based synthesizer to convert the description into a regex. 
\chm{If the synthesized regex is inconsistent with the given examples, it then leverages a regex repairer to modify the synthesized regex guided by the examples, and returns the revised regex.}

Considering that the lack of focus on \emph{validity}
in previous NLP-based works (e.g., on the dataset StructuredRegex, less than half of the regexes predicted by {\scshape Deep-Regex} (Locascio et al.)~\cite{DBLP:conf/emnlp/LocascioNDKB16} are valid, see Section~\ref{sec:rq1}), we propose a two-phase \chm{NLP-based synthesis} model \lyt{{\scshape $\rm S_2RE$}} to solve this problem \chm{as follows}. \chm{By rewarding the \lyt{{\scshape $\rm S_2RE$}} model with validity, in addition to the semantic correctness reward used in previous works \cite{DBLP:conf/naacl/KushmanB13,park-etal-2019-softregex}, our model is towards generating more valid regexes than previous models. If the generated regexes are still invalid after that, \iffalse it uses\fi the \lyt{invalid2valid} model, wherein the structure of \lyt{{\scshape $\rm S_2RE$}} is reused, is used to \iffalse , which  takes the invalid regexes as input and\fi transform them into valid ones to further guarantee the validity of regexes.}

There are various drawbacks or limitations in existing repair techniques, such as only \chm{supporting} positive \chm{or} negative examples~\cite{DBLP:conf/emnlp/LiKRVJ08,DBLP:conf/pakdd/RebeleTS18}, not \chm{supporting} regexes with the
conjunction ($\&$) operator, or \chm{having} the problem\lyt{s} of under-fitting/over-fitting~\cite{DBLP:journals/pacmpl/PanHXD19,ase20}.
 To overcome these drawbacks or limitations, we present a \chm{novel and} efficient algorithm, {\scshape SynCorr}, based on Neighborhood Search (NS) to repair an incorrect regex to achieve that the repaired regex is consistent with the examples. \chm{Particularly  {\scshape SynCorr}} \chm{alleviates} \chm{the} under-fitting/over-fitting \chm{problem}, via preserving the integrity of the small sub-regexes.
 
{\scshape TransRegex} can greatly reduce the aforementioned errors caused by ambiguity, imprecision or unknown words in  \lyt{NL} descriptions, via using the example-guided regex repairer. 
In comparison \chm{with existing multi-modal works~\cite{pldi20,tacl}}, {\scshape TransRegex} avoids \chm{their} limitations by not restricting the sketches of the generated regex. 
\chm{Furthermore, {\scshape TransRegex}} modularizes \chm{the synthesis problem as} the NLP-based regex synthesis and example-guided regex repair, allowing one to \chm{use his own algorithms or any other new algorithms instead}. 

We evaluate {\scshape TransRegex} by comparing {\scshape TransRegex} against ten state-of-the-art tools on three publicly available datasets.
Our evaluation demonstrates the accuracy of our {\scshape TransRegex}  is 17.4\%, 35.8\% and 38.9\% higher than that of NLP-based works on the three datasets, respectively.  Further, {\scshape TransRegex} can achieve higher accuracy than the state-of-the-art multi-modal works\chm{~\cite{pldi20,tacl}},  with 10\% to 30\% higher accuracy on all three datasets.
Our evaluation also reveals our NLP-based model \lyt{{\scshape $\rm S_2RE$}}  can generate 100\%  valid regexes on complex dataset, whereas other NLP-based tools can synthesize 49.6\% to 90.6\% valid ones. 
Finally, the evaluation results on regex repair also show that our {\scshape SynCorr} has better capability than existing repair tools. 


The contributions of this paper are listed as follow.
\begin{itemize}
    \item We propose {\scshape TransRegex}, an automatic framework which can synthesize regular expressions from both NL descriptions and examples. To the best of our knowledge, {\scshape TransRegex} is the first to treat the NLP-and-example-based regex synthesis problem as \chm{the problem of} NLP-based synthesis with regex repair. 
    \item We introduce a two-phase algorithm \lyt{{\scshape $\rm S_2RE$}} for regex synthesis from NL. By rewarding the \lyt{{\scshape $\rm S_2RE$}} model with validity and \chm{using} the \lyt{invalid2valid} model, \lyt{{\scshape $\rm S_2RE$}} generates more valid regexes \chm{while} having similar or higher accuracy than the state-of-the-art NLP-based models. 
    \item We present a novel algorithm {\scshape SynCorr} for regex repair that (\rmnum{1}) leverages Neighborhood Search (NS) algorithms to guide the search for a better regex which is consistent with the given examples from the neighborhoods of the incorrect regex, and (\rmnum{2}) utilizes some rewriting rules for \chm{sub-regexes} abstraction to preserve the integrity of some small sub-regexes, thereby \chm{alleviating} under-fitting/over-fitting and efficiently reducing the search space.
    \item We conduct a series of comprehensive experiments comparing {\scshape TransRegex} with ten state-of-the-art synthesis tools.
    The evaluation results demonstrate that the accuracy of our {\scshape TransRegex}  is 17.4\%, 35.8\% and 38.9\% higher than that of NLP-based approaches on the three datasets, respectively, while {\scshape TransRegex} can achieve higher accuracy than the state-of-the-art multi-modal techniques with 10\% to 30\% higher accuracy on all three datasets. The evaluation results also indicate {\scshape TransRegex} utilizing natural language and examples in a more effective way.
\end{itemize}

\section{Overview}\label{overview}
\begin{figure*}[htpb]
    \centering
    \includegraphics[width=0.8\linewidth]{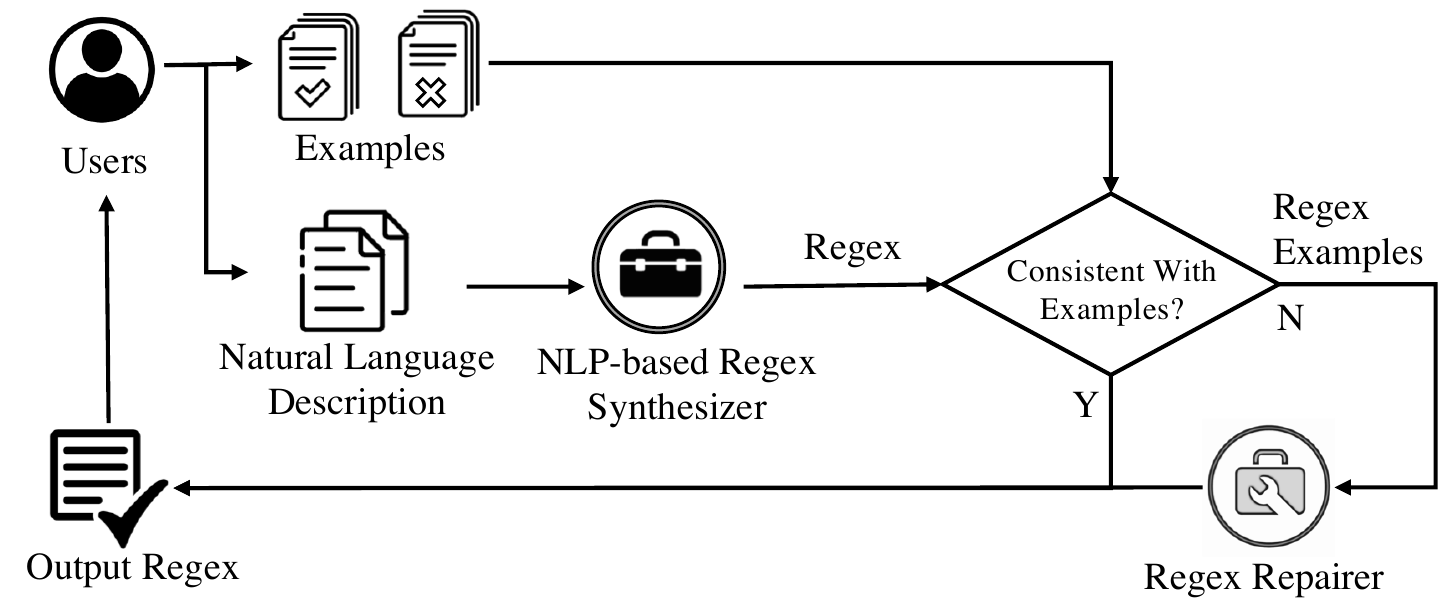}
    \caption{An overview of framework {\scshape TransRegex} for regex synthesis.}
    \label{fig:framework}
\end{figure*}

In this section, we present an overview of {\scshape TransRegex}.
As illustrated in Fig.~\ref{fig:framework}, {\scshape TransRegex} consists of two steps, namely, \chm{the} \emph{NLP-based regex synthesis} (Section~\ref{algsyn}) and \chm{the} \emph{example-guided regex repair} (Section~\ref{algrep}).
\chm{In} the \chm{first} step\chm{,} \emph{NLP-based regex synthesis} takes the given  \lyt{NL} description as input and tries to synthesize a regex from the  \lyt{NL} description via an NLP-based synthesizer. 
After that, if the synthesized regex is consistent with the given examples, \chm{then} {\scshape TransRegex} outputs the regex.
Otherwise, \emph{example-guided regex repair} modifies \chm{this synthesized} regex  based on the provided examples by an example-guided repairer, and returns the repaired regex. Next, we illustrate the main ideas behind {\scshape TransRegex} using a motivating example.


\begin{example}\label{example1}
Consider the task of \chm{constructing} a regex \verb|[AEIOUaeiou].*[0-9]{7,}.*|. \chm{The}  \lyt{NL} description $\mathcal{N}\mathcal{L}$, \chm{the} positive examples $\mathcal{P}$, and \chm{the} negative examples $\mathcal{N}$ \chm{are} shown in Fig.~\ref{fig:example1}.
\end{example}

\begin{figure}[!hbpt]
\centering

\scalebox{0.85}{
\begin{tikzpicture}
\node [draw] (example-tabular) {
\begin{tabular}{|c|c|}
  \hline
  \multicolumn{2}{|c|}{{\small Natural Language Description $\mathcal{N}\mathcal{L}$}}\\
  \hline
  \multicolumn{2}{|c|}{{\small{\it items with a vowel preceding a numeral \textbf{\textcolor{cRed}{\underline{at least 7 times}}}}}}\\
  \hline
  {\small Positive Examples $\mathcal{P}$}&{\small Negative Examples $\mathcal{N}$}\\
  \hline
{\small E18043699}&{\small u.}\\
{\small U530136382}&{\small jz;B}\\
{\small U65972791327}&{\small o45}\\
{\small U82433805}&{\small FBcW}\\
{\small i3390716928}&{\small I4k,S}\\
{\small O789821610}&{\small U}\\
{\small U4765749255}&{\small I\$\#].}\\
{\small E6204251}&{\small A}\\
{\small e6868266}&{\small uV}\\
{\small O50693106874}&{\small o20m3u5817}\\
  \hline
\end{tabular}
};
\end{tikzpicture}
}
\caption{A pair of a description  and examples.}\label{fig:example1}
\end{figure}

\begin{figure}[!hbpt]
\centering
\scalebox{0.9}{
\includegraphics[width=\linewidth]{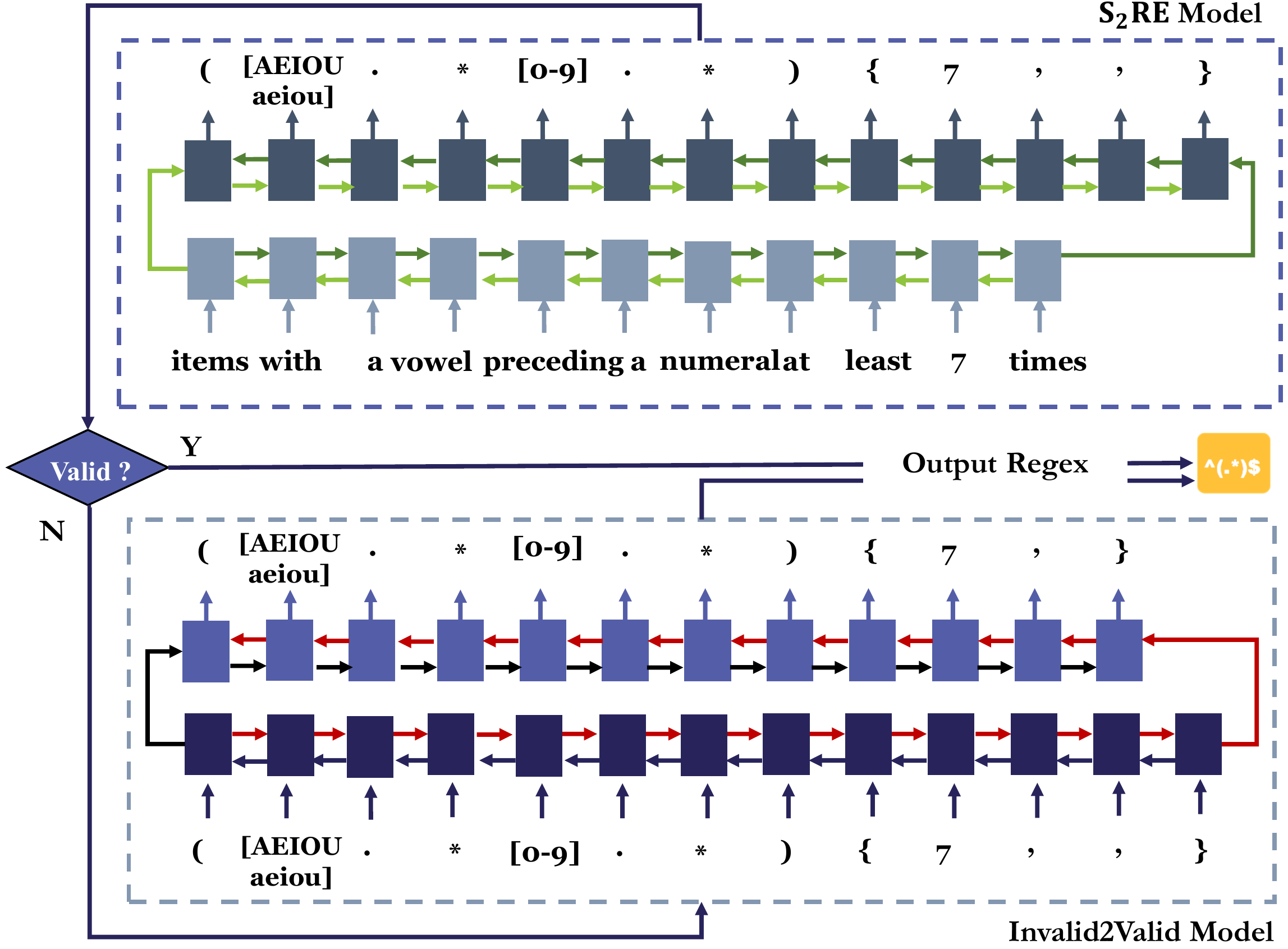}
}
\caption{The process of \chm{the} algorithm \lyt{{\scshape $\rm S_2RE$}}.}\label{fig:regex_synthesis_model_structure}
\end{figure}

First, {\scshape TransRegex} utilizes \chm{our} NLP-based synthesizer \lyt{{\scshape $\rm S_2RE$}} to translate \chm{the}  \lyt{NL} \chm{description $\mathcal{N}\mathcal{L}$} into a regex.  As shown in Fig.~\ref{fig:regex_synthesis_model_structure}, the encoder of \lyt{{\scshape $\rm S_2RE$}} generates latent vectors from the given description $\mathcal{N}\mathcal{L}$, and the decoder of \lyt{{\scshape $\rm S_2RE$}} synthesizes the corresponding regex
\verb|([AEIOUaeiou].*[0-9].*){7,,}| 
according to the latent vectors from the encoder. 
\chm{However,} it is clear that \chm{this} synthesized regex is invalid. \lyt{{\scshape $\rm S_2RE$}} \chm{then} converts \chm{this} invalid regex into a valid one \verb|([AEIOUaeiou].*[0-9].*){7,}| using our pretrained invalid2valid model.
It \chm{is easy to verify} that the \chm{valid} regex generated by  \lyt{{\scshape $\rm S_2RE$}} is inconsistent  with  the provided examples, such as \chm{the} positive example \verb|E18043699| $\notin L($\verb|([AEIOUaeiou].*[0-9].*){7,}|$)$.
Intrinsically the description $\mathcal{N}\mathcal{L}$ shown in Fig.~\ref{fig:example1} is ambiguous---\emph{i.e.},
it is unclear what part of the string \chm{should} appear at least 7 times, resulting in \chm{the incorrect regex generated by \lyt{{\scshape $\rm S_2RE$}}.} 

\begin{figure}
\centering
\scalebox{0.9}{
   \includegraphics[width=\linewidth]{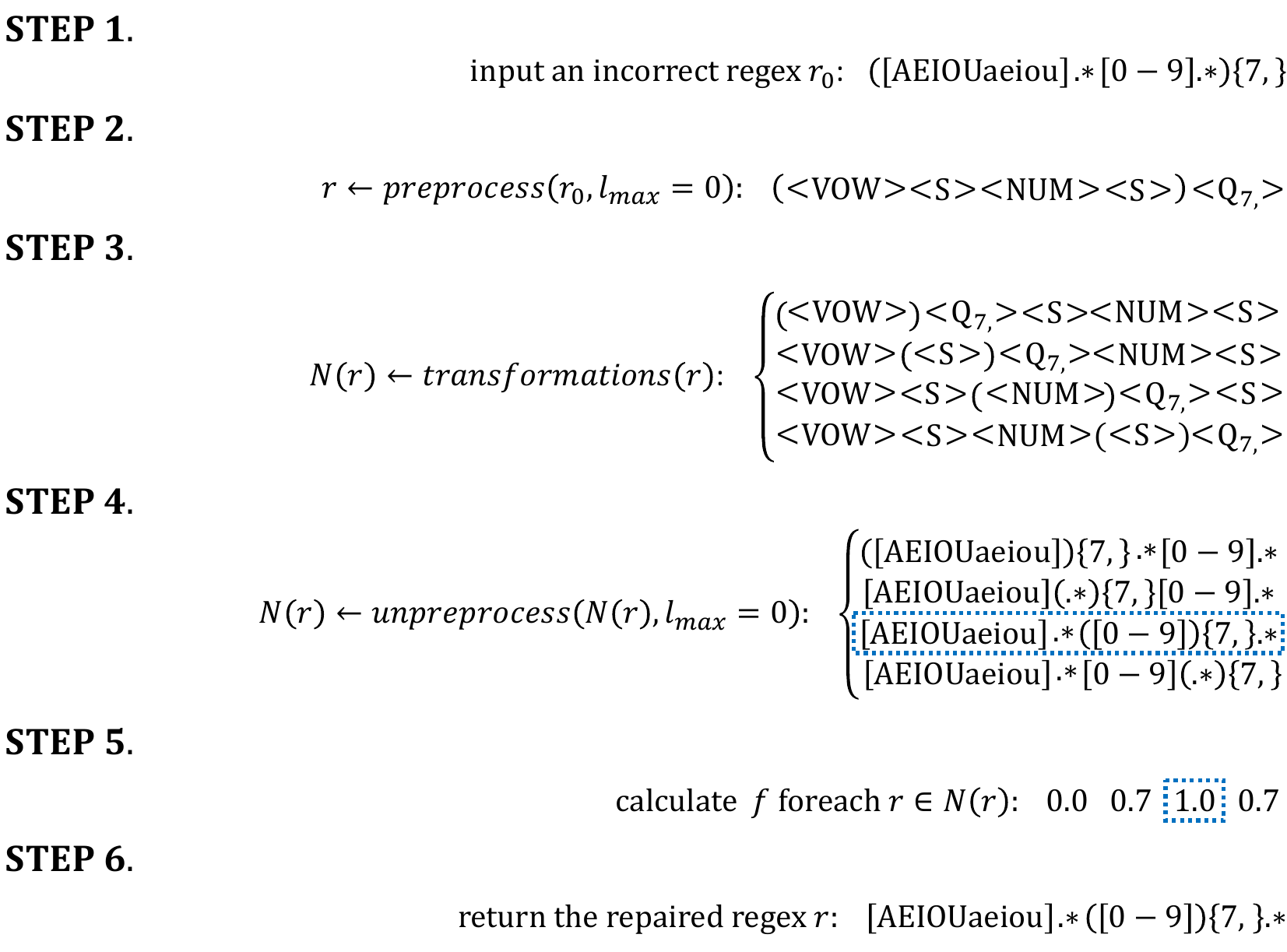}
}
\caption{The process of \chm{the} algorithm {\scshape SynCorr}.}\label{fig:exampleSynCorr} 

\end{figure}

\chm{So} after that, {\scshape TransRegex} employs \chm{the} algorithm {\scshape SynCorr} \chm{that is} \lyt{based on Neighborhood Search (NS)} to repair the incorrect regex, given in Fig.~\ref{fig:exampleSynCorr}.
As we \chm{have} mentioned in Section~\ref{sec1}, the incorrect regexes generated by \lyt{{\scshape $\rm S_2RE$}} may be very similar to the target regexes with subtle differences.
Therefore, {\scshape SynCorr} first converts  the above incorrect regex 
to the abstract regex $r=$ \verb|(<VOW><S><NUM><S>)<|\czx{${\rm Q_{7,}}$}\verb|>| with symbolic symbols \chm{obtained} by executing \chm{the} function \emph{\lyt{preprocess}},
so that the integrity of small sub-regexes (e.g., \verb|[AEIOUaeiou]| and \verb|[0-9]|) can be retained as much as possible in the subsequent steps. Then {\scshape SynCorr} calls \chm{the} function  \emph{\lyt{transformations}} 
to \lyt{get the neighbours of $r$, i.e.,}  some abstract regexes (e.g., \verb|<VOW><S>(<NUM>)<|\czx{${\rm Q_{7,}}$}\verb|><S>|) that are similar to \lyt{$r$} through a series of subtle transformations\footnote{\chm{In Fig.~\ref{fig:exampleSynCorr},} we only demonstrate transformation {\it quantifier adjustment} in \chm{STEP 3}.
For all transformations, see Section~\ref{sec:algorithm}.
} (e.g., 
{\it quantifier adjustment} or {\it element  replacement}, etc.). Next, {\scshape SynCorr} maps these abstract regexes into corresponding concrete regexes via using \chm{the} function \emph{\lyt{unpreprocess}}, and calculates the $f$ value\footnote{The $f$ value of a regex represents \chm{to which degree} \iffalse the degree to which\fi the regex meets \iffalse the consistency of\fi the given examples. Especially, \chm{a}  regex with $f$ value $1$ will accepts all positive examples $\mathcal{P}$ and rejects all negative examples $\mathcal{N}$.} of each regex. Finally, {\scshape SynCorr} returns the regex \verb|[AEIOUaeiou].*[0-9]{7,}.*| with $f$ value of $1$. Leveraging NLP-based synthesis with regexs repair, our {\scshape TransRegex} is able to synthesize
the correct one
mentioned above.
This success case shows that our {\scshape TransRegex} can well deal with the ambiguity of  \lyt{NL} and the \chm{in}validity of regexes produced by some NLP-based synthesizers.

In addition, it is worth noting that on the same example and the incorrect regex mentioned above,
the repair tool
{\scshape RFixer}~\cite{DBLP:journals/pacmpl/PanHXD19} produces the incorrect regex \verb|([AEeiO01234U56789].*[0-9].*){2,}|. 
Specifically,  {\scshape RFixer} cannot generalize for some unseen examples (e.g., a positive example \verb|a1234567| 
), this will result in the regex produced by {\scshape RFixer} without some characters like “a”, i.e., the generated regexes will be  over-fitting. \chm{In contrast}, {\scshape SynCorr} avoids over-fitting well by preserving the integrity of the small sub-regexes. 

\chm{However,} \lyt{considering that \iffalse our repair algorithm\fi {\scshape SynCorr}  is an algorithm based on NS \iffalse design\fi and \chm{thus} \iffalse it\fi may trap in local optimum, we will continue to use {\scshape RFixer} to repair if {\scshape SynCorr} fails.} \xzw{As demonstrated in TABLE~\ref{tab:rq2}, {\scshape SynCorr}+{\scshape RFixer} can  achieve  more than $10\%$ higher success rate of repair than {\scshape SynCorr} on the experimental datasets.} \chm{Further, if there will be more powerful repair tool \lyt{than} {\scshape RFixer}, by combining {\scshape SynCorr} we can achieve even higher success rate, which is a future work.}

\section{Regex Synthesis Algorithm}\label{sec:algorithm}
In this section, we present the details of our synthesis algorithm {\scshape TransRegex}.
Before \chm{that, we first provide the background.} 

\subsection{Background}\label{sec2}
Let $\Sigma$ be a finite alphabet of symbols. The set of all words over $\Sigma$ is denoted by $\Sigma^*$. The empty word and the empty set are denoted by $\varepsilon$ and $\varnothing$,  respectively.

\textbf{Regular Expression (Regex).}
{Expressions of $\varepsilon$, $\varnothing$, and $a \in \Sigma$ are regular expressions; a regular expression is also formed using the operators
$$[C] \quad r_1 | r_2\quad  r_1r_2 \quad r_1\&r_2 \quad  (r) \quad {\footnotesize \sim} r_1 \quad r_1\{m,n\}$$
where $C \subseteq \Sigma,~C\neq \{\varepsilon\}$ or $\varnothing$ is a set of characters, $m\in\mathbb{N}$, $n\in\mathbb{N}\cup\{{\infty\}}$, and $m\leq n$.}
Besides, $r?$, $r^*$, $r^+$ and $r\{i\}$ where $i\in\mathbb{N}$ are abbreviations of $r\{0,1\}$, $r\{0,\infty\}$, $r\{1,\infty\}$ and $r\{i,i\}$, respectively. $r\{m,\infty\}$ is often simplified as $r\{m,\}$.
The {\it language} $L(r)$ of a regular expression $r$ is defined inductively as follows:
$L(\varnothing)=\varnothing$;
$L(\varepsilon)=\{\varepsilon\}$;
$L(a)=\{a\}$;
$L([C])=C$;
$L(r_1|r_2)=L(r_1)\cup L(r_2)$;
$L(r_1 r_2)= \{vw~|~v\in L(r_1), w\in L(r_2)\}$;
$L(r_1\&r_2)= \{v~|~v\in L(r_1)\wedge v\in L(r_2)\}$;
$L({\footnotesize \sim} r_1)= \{v~|~v\notin L(r_1)\}$;
$L(r\{m,n\})=\bigcup_{m\leqslant i \leqslant n} L(r)^{i}$.

If \chm{an} expression follows the syntax above, then \chm{it} is a\emph{ valid} \chm{one}, otherwise it is\emph{ invalid}. For instance, the expression \verb|ab{1,3}| is valid, but the expression \verb|ab{1,,,,3}| is invalid.

\subsection{The Main Algorithm}

Our synthesis algorithm is shown in Algorithm~\ref{al1}, which aims to synthesize regexes from \lyt{NL} descriptions and examples.
In detail, our algorithm first employs a\lyt{n} NLP-based synthesizer \lyt{{\scshape $\rm S_2RE$}} to generate a regex $r$ from the given \lyt{NL} description $\mathcal{N}\mathcal{L}$ (line $1$), which is introduced in Section~\ref{algsyn}.
Then, if $r$ is consistent with the given positive and negative examples, {\scshape TransRegex} outputs \chm{$r$} (line $2$).
Otherwise, {\scshape TransRegex} leverages two example-guided repairers to fix the incorrect regex \chm{$r$} based on the provided positive and negative examples, which are described in Section~\ref{algrep}, and returns the repaired regex (lines $3$--$6$).

\begin{algorithm}[htp]
  \caption{{\scshape TransRegex}}\label{al1}
	\LinesNumbered
    \KwIn{a natural language description $\mathcal{N}\mathcal{L}$, positive examples $\mathcal{P}$, and negative examples $\mathcal{N}$}
    \KwOut{a regex}

    $r \leftarrow$ \lyt{{\scshape $\rm S_2RE$}}($\mathcal{N}\mathcal{L}$);\\

     \lIf{$\mathcal{P}\subseteq L(r)$ and $ \mathcal{N} \cap L(r) = \varnothing$}{
     \textbf{return} $r$
     }
    \Else{
    $r\leftarrow $ {\scshape SynCorr}($r$);\\
       \lIf{$\mathcal{P}\subseteq L(r)$ and $ \mathcal{N} \cap L(r) = \varnothing$}{
     \textbf{return} $r$
     }
       \lElse{\textbf{return} {\scshape RFixer}($r$)}
    }
\end{algorithm}

\subsection{Regex Synthesis from Natural Language \chm{Descriptions}}\label{algsyn}

To synthesize a regex from the \lyt{NL} description, we build a seq2seq model with attention mechanism as our regex synthesis model \lyt{{\scshape $\rm S_2RE$}}. It consists of an encoder and a decoder. The encoder initializes the words in the \lyt{NL} sequences as vectors, then it encodes the vectors as hidden states, which represents the semantic representations of the \lyt{NL} sequences. The decoder generates the corresponding regex according to the latent representations from the encoder.

The training of our neural \lyt{{\scshape $\rm S_2RE$}} model consists of two stages, using two different strategies.

\textbf{Maximum Likelihood Estimation (MLE)}: In the first stage, we use MLE to maximize the likelihood of mapping the \lyt{NL} description to corresponding regex:
\begin{equation}
  \theta={arg max}\sum_{(\mathcal{N}\mathcal{L},r)\in D}log(p(r|\mathcal{N}\mathcal{L}))
\end{equation}
where $D$ is the training set, $p(r|\mathcal{N}\mathcal{L})$ is the probability that the seq2seq model generates a regex $r$ from a \lyt{NL} description $\mathcal{N}\mathcal{L}$.

\textbf{Policy Gradient}: MLE may fail to consider the semantic equivalence of the regexes that might be different in syntax and the validity of the regexes (especially for complex and realistic datasets). Therefore, in the second stage, we gradually train our \lyt{{\scshape $\rm S_2RE$}} model via policy gradient~\cite{policygradient} by rewarding the model according to the following two indicators.
\begin{itemize}
    \item Semantic Correctness: following Park et al.'s work~\cite{park-etal-2019-softregex}, we reward the \lyt{{\scshape $\rm S_2RE$}} model if it generates \chm{a} regex that is semantically equivalent to the ground truth, that is, the semantic reward $R_C(r)$ is $1$ if the regex r is semantically equal with ground truth and 0 otherwise;
    \item Syntactic Validity: we also reward the \lyt{{\scshape $\rm S_2RE$}} model if it generates regexes that are valid\footnote{The syntax-checking is implemented via the re.compile() function in Python.}, that is, the syntactic reward $R_V(r)$ is $1$ if the regex r is valid and 0 otherwise.
\end{itemize}

Finally, the objective of the second stage is to maximize the following function:
\begin{equation}
  J(\theta)=\sum_{(\mathcal{N}\mathcal{L},r)\in D}p(r|\mathcal{N}\mathcal{L})R(r)
\end{equation}

\begin{equation}
      R(r)=\alpha R_C(r) + \beta R_V(r)
\end{equation}
where $\alpha$ and $\beta$ are hyper-parameters.

The \lyt{{\scshape $\rm S_2RE$}} model helps to generate more correct and valid regexes than the previous models. However, it still can not guarantee to generate valid regexes for all the input \lyt{NL} descriptions. To solve this problem, we reuse the structure of the \lyt{{\scshape $\rm S_2RE$}} model to build our invalid2valid model (wherein only $R_V$ is used). Specifically, the invalid2valid model is trained on $5,000$ invalid and valid regex pairs, which are collected as following:
\begin{itemize}
    \item Get a valid regex randomly from the dataset \lyt{{\it StructuredRegex}~\cite{DBLP:conf/acl/YeCDD20}};
    \item \chm{Make it invalid by} performing some minor changes on it : adding or deleting or modifying 1-5 positions randomly;
    \item If the changed regex is still valid, then discard it.
\end{itemize}


The whole process of generating regexes from \lyt{NL} descriptions is shown in Fig.~\ref{fig:regex_synthesis_model_structure}. \chm{It shows that, the algorithm first uses our \lyt{{\scshape $\rm S_2RE$}} model to generate a regex. If the generated regex is invalid, it then transform this regex fast to a valid one using the pretrained invalid2valid model.} The experiments show that after the \lyt{{\scshape $\rm S_2RE$}} model and the invalid2valid model, we obtain $100\%$ valid regexes in syntax.

\subsection{Regex Repair from Examples}\label{algrep}
The regexes generated by \lyt{{\scshape $\rm S_2RE$}} may be incorrect but very similar to the target regexes with subtle differences.
In this section, we present our algorithm \lyt{{\scshape SynCorr}} to repair these incorrect regexes.
The key idea is to search for a better regex which accepts \lyt{more} positive examples and rejects \lyt{more} negative examples from the neighborhoods of input regex.

\czx{To start with, we define the \emph{neighborhood} and an evaluation criterion of regex $r$. Given a regex $r$, we define its \emph{neighborhood}, denoted as $N(r)$, as the set of regexes, which can be obtained by applying a transformation on $r$ \chm{(transformations are given later)}. 
In order to select a regex $r$ among a set of regexes, we define a measure $f$ on $r$ with respects to positive examples $\mathcal{P}$ and negative examples $\mathcal{N}$ as
\begin{equation}
  f(r, \mathcal{P}, \mathcal{N}) = \frac{|T_{\mathcal{P}}| + |T_{\mathcal{N}}| - |F_{\mathcal{P}}| - |F_{\mathcal{N}}|}{|\mathcal{P}| + |\mathcal{N}|}
\end{equation}
where $T_{\mathcal{P}} = \{w \in L(r)|w \in \mathcal{P}\}$, $T_{\mathcal{N}} = \{w \notin L(r)|w \chm{\in} \mathcal{N}\}$, $F_{\mathcal{P}} = \{w \notin L(r)|w \in \mathcal{P}\}$, $F_{\mathcal{N}} = \{w \in L(r)| w \chm{\in}  \mathcal{N}\}$.
Intuitively, the higher the $f$ value, the better the regex $r$. Especially, the regex $r$ with $f$ value of $1$ will accepts all positive examples and rejects all negative examples.
}




\begin{algorithm}[htp]
  \caption{{\lyt{{\scshape SynCorr}}}}\label{alNS}
  \LinesNumbered
    \KwIn{positive examples $\mathcal{P}$, negative examples $\mathcal{N}$, and \lyt{an incorrect} regex $r_0$}
    \KwOut{a regex}
    \For(){\lyt{$l_{max}$} {\rm = 2 to 0}}{
      $r \leftarrow \lyt{preprocess(r_0, l_{max})}$;\\
      \While{${\rm the}\ stop\ conditions\ {\rm not\ meet}$}{
        $N(r) \leftarrow$ apply the transformations on $r$ \textbf{in parallel};\\
        \lyt{$r \leftarrow unpreprocess(r, l_{max})$};\\
         \lyt{$N(r) \leftarrow unpreprocess(N(r), l_{max})$};\\
        $r_m \leftarrow \mathop{\arg\max}\limits_{r' \in N(r)} f(r', \mathcal{P}, \mathcal{N})$;\\


        \If{$f(r_m, \mathcal{P}, \mathcal{N}) > f(r, \mathcal{P}, \mathcal{N})$}{
          \lIf{$f(r_m, \mathcal{P}, \mathcal{N}) == 1$}{
            \textbf{return} $r_m$
          }
          \lElse(){$r\leftarrow \lyt{preprocess(r_{m}, l_{max})}$}
        }
        \lElse(){\textbf{break}}
      }
    }      
    \textbf{return} $r_0$;
\end{algorithm}

\czx{
The \lyt{algorithm {\scshape SynCorr}} is shown in Algorithm \ref{alNS}.
In order to facilitate neighborhood search, \lyt{{\scshape SynCorr}} sets 
the highest level of rewriting rules $l_{max}$ ranges from $2$ to $0$ (line $1$). 
First, \lyt{{\scshape SynCorr}} preprocesses the given regex $r_0$ with the given highest level $l_{max}$
and keeps the result regex in $r$ (\lyt{line $2$}).
Next, it applies the transforms on $r$ to get the neighborhood $N(r)$ and unpreprocess $r$ and $N(r)$ (\lyt{lines \chm{$4$--$6$}}).
Then \lyt{{\scshape SynCorr}} selects the regex with the maximum $f$ value, denoted as $r_{m}$, from the neighborhoods of $r$ (\lyt{line $7$}). After that, it compares the $f$ values between the current regex $r$ and the selected regex $r_m$ (\lyt{line $8$}).
If the selected regex $r_m$ gets a higher $f$ value, then \lyt{{\scshape SynCorr}} checks whether the $f$ value equals to $1$. If it is, \lyt{{\scshape SynCorr}} returns $r_m$ (\lyt{line $9$}). Otherwise, to start with the next iteration, the regex $r_{m}$ is preprocessed and assigned to $r$ (\lyt{line $10$}).
If the selected regex $r_m$ does not get a higher $f$ value, then $r_m$ may be a local maximum, so \lyt{{\scshape SynCorr}} fails to repair regex $r_0$ and breaks the loop (\lyt{line $11$}).
For each $l_{max}$, the processing above runs until the stop conditions meet (\lyt{lines \chm{$3$--$11$}}). 
Finally, \lyt{{\scshape SynCorr}} returns $r_0$ if \lyt{{\scshape SynCorr}} fails for all $l_{max}$ values (\lyt{line $12$}).
}

\textbf{Rewriting rules.}
In order to preserve the integrity of the small sub-regexes (probably the correct part) and reduce the search space, we define some rewriting rules for the regexes to abstract some small sub-regexes. \chm{The resulting regexes are called the abstract forms of the original regexes.} Specifically, we rewrite some special small regexes to unique symbolic nodes, which are listed as follows:

\hbox to \hsize{\vbox{\hsize=0.5\hsize  $$
\begin{array}{rl}
[C]&\rightarrow_{0} { <C}_{C}>\\
\{m,n\}&\rightarrow_{0} { <Q}_{m,n}>\\
.* r &\rightarrow_{1} { <SL}_{r}>\\
.* r.* &\rightarrow_{1} { <SLR}_{r}>\\
{\footnotesize \sim} (r.* ) &\rightarrow_{2}{ <NSR}_{r}>\\
\end{array}
$$}\vbox{\hsize=0.5\hsize $$
\begin{array}{rl}
const &\rightarrow_{0} { <C}_{const}>\\
{\footnotesize \sim} (r) &\rightarrow_{1} {<N}_{r}> \\
r.* &\rightarrow_{1} {<SR}_{r}>\\
{\footnotesize \sim} (.* r ) &\rightarrow_{2} {<NSL}_{r}>\\
{\footnotesize \sim} (.* r.*) &\rightarrow_{2} {<NSLR}_{r}>\\
\end{array}
$$}}
where $const$ denotes a string (in regex) consisting of characters in $\Sigma$ (\emph{e.g.}, ``abc''), $r$ is $[C]$ or a $const$, and the suffix \czx{indicates the level $l$ of the rule}.
The rewriting rules are performed greedily and depending on \czx{its level}. In our implementation, we use some meaningful names for some special regexes (\emph{e.g}, use \verb|<NUM>|, \verb|<LET>|, \verb|<CAP>|,   and \verb|<VOW>| for \verb|[0-9]|, \verb|[A-Za-z]|, \verb|[A-Z]|,  and \verb|[AEIOUaeiou]|,  respectively) and keeps the rewriting mappings in dictionaries (\emph{i.e.}, \czx{$Dict_{l}$} for the \czx{level $l$}).

\textbf{Preprocess and unpreprocess.}
Given a regex $r$
\czx{and the highest level $l_{max}$}, the preprocess is to abstract (\emph{i.e.}, from left to right) $r$ according the rewriting rules whose \czx{level $l$} ranging from \czx{$l_{max}$} to 0 in order.
The rules with high level are performed first.
Take $r_0$ in Fig.~\ref{fig:exampleSynCorr} \chm{as an} example.
If \czx{$l_{max} = 2$}, the rules with level $l = 2$ are performed first, but without changing $r_0$ because of no rules with level $l = 2$ are applicable. Then the rules with level $l = 1$ are \chm{applied}, yielding $r_1 = $\verb|(<SRVOW><SRNUM>){7,}|. Finally, the rules with level $l = 0$ is applied, yielding the final \czx{preprocessed} regex $r = $\verb|(<SRVOW><SRNUM>)<|${\rm Q_{7,}}$\verb|>|.
If $l_{max} = 1$, the same preprocessed regex $r$ is returned. If $l_{max} = 0$, directly apply the rewriting rules with level $l=0$, yielding the final preprocessed regex \verb|(<VOW><S><NUM><S>)<|${\rm Q_{7,}}$\verb|>|.
The unpreprocess is the reversion of the preprocess.

\czx{\textbf{Stop conditions.} We can set the maximum number of iterations, and the maximum running time of the program as independent or mixed stop conditions.}

\textbf{Transformation.}
We  observe  that  most  of  the  incorrect regexes  generated by \lyt{{\scshape $\rm S_2RE$}}  can  be  made  equivalent to the  target regexes  with  only  minor  modifications.
For that, we design a series of transformations on regexes, \chm{which are listed bellow, where}
we use \emph{element}s to denote the small regexes that the rewriting rules can apply on, and \emph{generalized element}s to denote the regexes containing at least a pair of brackets.
\begin{itemize}
  \item {\it Binary Element Insertion}: this transformation inserts a binary element (\emph{i.e.}, disjunction, concatenation, or conjunction) from a candidate set into \lyt{the current} regex.
  \item \czx{{\it(Generalized) Element Deletion}: this transformation deletes a (generalized) element from \lyt{the current} regex.}
  \item \czx{{\it(Generalized) Element Replacement}}: this transformation replaces a \czx{(generalized)} element in \lyt{the current} regex with an element from a candidate set.
  \item {\it Quantifier Insertion}: this transformation inserts a quantifier, whose minimum and maximum \chm{values} are selected according to the positive and negative examples, to a (generalized) element in \lyt{the current} regex.
  \item {\it Quantifier Modification}: this transformation modifies a quantifier of a (generalized) element in \lyt{the current} regex, where the minimum and maximum \chm{values} are set according to the positive and negative examples.
  \item \czx{{\it Quantifier Adjustment}: this transformation adjusts the (generalized) element restricted by a quantifier in \lyt{the current} regex from its original restrict (generalized) element to another (generalized) element.}
  \item \czx{{\it Operator Insertion}: this transformation inserts an operator (\emph{i.e.}, negation, disjunction, or conjunction) into \lyt{the current} regex.}
  \item \czx{{\it Operator Deletion}: this transformation deletes an operator from \lyt{the current} regex.}
  \item {\it Element Adjustment}: this transformation adjusts an element in \lyt{the current} regex from its original position to another position.
  \item {\it(Generalized) Element Exchanging}: this transformation swaps two (generalized) elements in \lyt{the current} regex.
\end{itemize}
In our implementation, we take the elements collected in the dictionaries as the candidate set. And to search the neighbour fast, we perform the transformation in parallel, as there are no data races between each transformation. 

\section{Evaluation}


We implemented {\scshape TransRegex} in Python, and conducted experiments on a machine with 16 cores Intel Xeon CPU E5620 @ 2.40GHz with 12MB Cache, 24GB RAM, running Windows 10 operating system.
Under this experiment settings, we then designed our experiments to answer 
the following research questions:

\begin{itemize}
   \item  RQ1: Can  \lyt{{\scshape $\rm S_2RE$}} model generate correct and valid regexes from natural language descriptions? (\S\ref{sec:rq1})
    \item RQ2: Can {\scshape SynCorr} repair incorrect regexes from examples? (\S\ref{sec:rq2})
    \item RQ3: Can \textsc{TransRegex} synthesize  regexes \chm{accurately}? (\S\ref{sec:rq3})
   
   \item RQ4:  Can {\scshape TransRegex} synthesize regexes efficiently? (\S\ref{sec:rq4})
\end{itemize}

\subsection{Datasets}
In the experiment, we evaluate {\scshape TransRegex} on three public datasets: {\it KB13}~\cite{DBLP:conf/naacl/KushmanB13}, {\it NL-RX-Turk}~\cite{DBLP:conf/emnlp/LocascioNDKB16}, and {\it StructuredRegex}~\cite{DBLP:conf/acl/YeCDD20}. \lyt{Among them, KB13 consists of 824 pairs of \lyt{NL} descriptions and the corresponding regexes  constructed by regex experts.
NL-RX-Turk includes 10,000 pairs of \lyt{NL} descriptions and regexes collected through crowdsourcing.
StructuredRegex comprises 3,520 long English descriptions which are 2.9 to 4.0 times longer than the first two datasets, paired with complex regexes and associated 6 positive/6 negative examples using crowdsourcing.
As our approach requires examples which are absent in the first two datasets,  we adopt the corresponding 10 positive and 10 negative examples for the first two datasets provided by Ye et al.~\cite{tacl}.
Specifically, the 10 positive examples are enumerated by randomly traversing the deterministic finite automaton (DFA) of the given regex (resp. the 10 negative examples are synthesized by stochastically traversing the DFA of the negation of the given regex).
}

\subsection{Training Setting}

\lyt{We train {\scshape $\rm S_2RE$} on the three public datasets shown in TABLE~\ref{tab:data}. We adopt the same train/validation/test sets as those used by previous works for the sake of comparison.}

\begin{table}[thbp]
  \caption{The Train, Validation, and Test Sets in Three Datasets.}
  \label{tab:data}
  \renewcommand\arraystretch{1.2}
\begin{adjustbox}{width=1.0\linewidth,center}
\begin{tabular}{|l|c|c|c|c|c|c}
    \hline
    \textbf{Dataset} &\textbf{Train Set}&\textbf{Validation Set}&\textbf{Test Set} \\
    \hline
    \hline
    {\small KB13} &618& 206 &206\\
    \hline
    {\small NL-RX-Turk}&6500&1000&2500\\
    \hline
    {\small StructuredRegex}&2173&351&996\\
    \hline
\end{tabular}
\end{adjustbox}
\end{table}

\lyt{In detail, the encoder and the decoder of the {\scshape $\rm S_2RE$} model consist of two stacked BiLSTM layers, respectively. The dimension size of the word embeddings is set to 128 and the hidden size is 256. We train the {\scshape $\rm S_2RE$} model with MLE loss for 30, 10 and 15 epochs on KB13, NL-RX-Turk and StructuredRegex, respectively. Then we further train the model with policy gradient for 10, 30,10 epochs on KB13, NL-RX-Turk and StructuredRegex, respectively.  The hyper-parameters $\alpha$ and $\beta$ are set to 0.5 and 0.5.}

\subsection{Baselines}
We compare three variants of {\scshape TransRegex} (i.e., {\scshape TransRegex} (\lyt{{\scshape $\rm S_2RE$}} + {\scshape SynCorr}), {\scshape TransRegex} (\lyt{{\scshape $\rm S_2RE$}} + {\scshape RFixer}), {\scshape TransRegex} (\lyt{{\scshape $\rm S_2RE$}} + {\scshape SynCorr} + {\scshape RFixer})) with ten relevant works, including our NLP-based algorithm \lyt{{\scshape $\rm S_2RE$}}. They are mainly fall into two paradigms. NLP-based works only used \lyt{NL} descriptions to synthesize regexes, i.e., {\scshape Semantic-unify}~\cite{DBLP:conf/naacl/KushmanB13}, {\scshape Deep-Regex} (Locascio et al.)~\cite{DBLP:conf/emnlp/LocascioNDKB16}, {\scshape Deep-Regex} (Ye et al.)~\cite{tacl}, {\scshape SemRegex}~\cite{DBLP:conf/emnlp/ZhongGYPXLLZ18}, {\scshape SoftRegex}~\cite{park-etal-2019-softregex}, and our \lyt{{\scshape $\rm S_2RE$}}. On the other hand, NLP-and-example-based works took both \lyt{NL} and examples for regex synthesis, including {\scshape Deep-Regex} (Ye et al.) + {\scshape Exs}~\cite{tacl}, {\scshape GrammarSketch}+ MLE~\cite{tacl}, {\scshape DeepSketch} + MLE~\cite{tacl} and {\scshape DeepSketch} + MML~\cite{tacl}.
Among them, {\scshape Semantic-unify} learns a probabilistic grammar model to parse \lyt{NL} descriptions into regexes. 
{\scshape Deep-Regex} (Locascio et al.), {\scshape Deep-Regex} (Ye et al.), {\scshape SemRegex}, {\scshape SoftRegex}, and our \lyt{{\scshape $\rm S_2RE$}}  are a series of algorithms and the main idea of these algorithms is to translate \lyt{NL} into regexes based on the seq2seq model.
{\scshape Deep-Regex} (Ye et al.) + {\scshape Exs} 
is an extension of {\scshape Deep-Regex} (Ye et al.), which takes examples into account by simply filtering the $k$-best regexes based on whether regexes consistent with the  given examples.
In addition, {\scshape GrammarSketch}+ MLE, {\scshape DeepSketch} + MLE, and {\scshape DeepSketch} + MML are a family of algorithms and these algorithms first use a grammar-based or
neural semantic parser to parse the \lyt{NL} 
into sketches, then search the regex space defined by the sketches and find a regex that is consistent with the given examples.

For implementation, {\scshape Deep-Regex} (Locascio et al.)~\cite{DBLP:conf/emnlp/LocascioNDKB16} and {\scshape SoftRegex}~\cite{park-etal-2019-softregex} are open-source programs, so we are able to reproduce the results of them. While other baselines do not release their source code, so we excerpted the statistics from their paper, and left blanks if they did not report it.





   




\subsection{RQ1: Effectiveness of  \lyt{{ $ S_2RE$}}}\label{sec:rq1}

To answer the first question, we compare our algorithm \lyt{{\scshape $\rm S_2RE$}} with five state-of-the-art NLP-based algorithms.
TABLE~\ref{tab:accuracy} shows the evaluation results on accuracy of {\scshape Semantic-Unify}, {\scshape Deep-Regex} (Locascio et al.), {\scshape Deep-Regex} (Ye et al.), {\scshape SemRegex}, 
{\scshape SoftRegex}, and our model \lyt{{\scshape $\rm S_2RE$}}.
On these three datasets, the accuracy of our \lyt{{\scshape $\rm S_2RE$}} model is similar to or slightly better than {\scshape SoftRegex}, and is always better than the other four NLP-based models (i.e., {\scshape Semantic-Unify}, {\scshape Deep-Regex} (Locascio et al.), {\scshape Deep-Regex} (Ye et al.), and {\scshape SemRegex}).

Besides, the evaluation results on validity are summarized in TABLE~\ref{tab:valid}.
The validity of synthesized regexes is crucial.
It guarantees the quality of regex synthesis from \lyt{NL} . \chm{Further, in  our approach, it} ensures that  the regex synthesized from \lyt{NL} is provided as a valid input to \emph{example-guided regex repair}. 
The results show that the validity of existing tools  (e.g., {\scshape Deep-Regex} (Locascio et al.), and {\scshape SoftRegex}) is unsatisfactory  on the last dataset StructuredRegex which is much more complex than the first two.
As we can see, less than half of the regexes generated by 
{\scshape Deep-Regex} (Locascio et al.) are valid and the most advanced NLP-based model {\scshape SoftRegex} achieves 90.6\%.
By contrast, our model \lyt{{\scshape $\rm S_2RE$}} achieves 100\% validity ratio, because it utilizes both the syntactic validity reward and invalid2valid model.
 \begin{table}[thbp]
  \caption{The DFA-equivalent Accuracy on Three Datasets.}
  \label{tab:accuracy}
    \renewcommand\arraystretch{1.2}
\begin{adjustbox}{width=1.0\linewidth,center}
\begin{tabular}{|l|c|c|c|c|c|c}
    \hline
    \multirow{2}{*}{\textbf{Approach}} &\multirow{2}{*}{\textbf{KB13}}&\multirow{2}{*}{\textbf{NL-RX-Turk}}&\textbf{Structured} \\
    &&&\textbf{Regex}\\
    \hline
    \hline
    {\small{\scshape Semantic-Unify}} &65.5\% &38.6\% &1.8\%\\
    {\small{\scshape Deep-Regex} (Locascio et al.)} &65.6\%& 58.2\% &23.6\%\\
    {\small{\scshape Deep-Regex} (Ye et al.)} &66.5\%& 60.2\%& 24.5\%\\
    {\small{\scshape SemRegex}}&78.2\%&62.3\%&--- \\
    {\small{\scshape SoftRegex}}&78.2\%&62.8\%&28.2\%\\
    {\small\lyt{{\scshape $\rm S_2RE$}}}&78.2\%&62.8\%&28.5\%\\
    \hline
    \hline
     {\small{\scshape Deep-Regex} (Ye et al.) + {\scshape Exs}} &77.7\%&83.8\%& 37.2\%\\
     {\small{\scshape GrammarSketch}+ MLE}&68.9\%&69.6\%&---\\
     {\small{\scshape DeepSketch} + MLE}&84.0\%&85.2\%&---\\
     {\small{\scshape DeepSketch} + MML}&86.4\%&84.8\%&---\\
    \hline
    \hline

     {\small{\scshape TransRegex} (\lyt{{\scshape $\rm S_2RE$}} + {\scshape SynCorr})}&92.7\%&94.2\%&63.3\%\\
    {\small{\scshape TransRegex} (\lyt{{\scshape $\rm S_2RE$}} + {\scshape RFixer})}&90.3\%&94.0\%&53.1\%\\
     {\small{\scshape TransRegex} (\lyt{{\scshape $\rm S_2RE$}} + {\scshape SynCorr} + {\scshape RFixer})}&\textbf{95.6\%}&\textbf{98.6\%}&\textbf{67.4\%}\\
 \hline
\end{tabular}
\end{adjustbox}
\end{table}

\begin{table}[thbp]
  \caption{The Number of Valid Regexes Generated by the Three NLP-based Models on Three Datasets.}
  \label{tab:valid}
  \renewcommand\arraystretch{1.2}
\begin{adjustbox}{width=1.0\linewidth,center}
\begin{tabular}{|l|c|c|c|c|c|c}
    \hline
    \multirow{2}{*}{\textbf{Approach}} &\multirow{2}{*}{\textbf{KB13}}&\multirow{2}{*}{\textbf{NL-RX-Turk}}&\textbf{Structured} \\
    &&&\textbf{Regex}\\
    \hline
    \hline
    {\small{\scshape Deep-Regex} (Locascio et al.)} &205 (99.5\%)& \textbf{2500 (100\%)} &494 (49.6\%)\\
    \hline
    {\small{\scshape SoftRegex}}&204 (99.1\%)&\textbf{2500 (100\%)}&902 (90.6\%)\\
    \hline
    {\small\lyt{{\scshape $\rm S_2RE$}}}&\textbf{206 (100\%)}&\textbf{2500 (100\%)}&\textbf{996 (100\%)}\\
    \hline
\end{tabular}
\end{adjustbox}
\end{table}

 \begin{mdframed}[style=MyFrame]
\textbf{Summary to RQ1:} \lyt{{\scshape $\rm S_2RE$}} can achieve similar or better accuracy  than the state-of-the-art NLP-based models. Meanwhile,  \lyt{{\scshape $\rm S_2RE$}} can synthesize more valid regexes. The advantage of high validty of \lyt{{\scshape $\rm S_2RE$}} becomes more obvious on complex datatsets.
\end{mdframed}

\subsection{RQ2: Effectiveness of SynCorr}\label{sec:rq2}
To answer the second question, \chm{we collected $45$, $930$ and $712$ incorrect regexes predicted by \lyt{{\scshape $\rm S_2RE$}}  on KB13, NL-RX-Turk, and StructuredRegex, respectively.} We compared {\scshape SynCorr} with the state-of-the-art tool {\scshape RFixer}.
TABLE~\ref{tab:rq2} shows the number of successful repairs\footnote{The regex that is successfully repaired must not only consistent with the given examples, but also be equal to the target regex.}.
{\scshape RFixer} successfully repaired 55.6\%, 83.9\%, and 34.4\% on dataset KB13, NL-RX-Turk, and StructuredRegex respectively.
In contrast, {\scshape SynCorr} can repair 11.1\%, 0.5\%, and 14.2\% more regexes than {\scshape RFixer} on the three datasets, respectively. 
In addition, the repair success rate of  {\scshape SynCorr} + {\scshape RFixer} is 11.1\% (22.2\%), 11.8\% (12.3\%), and 5.8\% (20.0\%) higher than that of {\scshape SynCorr} ({\scshape RFixer}) alone on dataset KB13, NL-RX-Turk, and StructuredRegex respectively.

We further demonstrate the advantages and disadvantages of {\scshape RFixer}  and {\scshape SynCorr} through a few cases in TABLE~\ref{tab:repair}.
As cases \#1 and \#2  shown in TABLE~\ref{tab:repair}, our algorithm {\scshape SynCorr} has the high  generalization performance compared with  {\scshape RFixer}. Case \#3  in TABLE~\ref{tab:repair} illustrates that
{\scshape SynCorr} can handle regexes with \& well but {\scshape RFixer} cannot. Similar to case \#4 in TABLE~\ref{tab:repair}, in some cases, {\scshape SynCorr} will fall into a local optimum, and {\scshape RFixer} can solve them.
Based on the above analysis, we can conclude that {\scshape SynCorr} and {\scshape RFixer} are complementary and simultaneously useful for our {\scshape TransRegex}.

 \begin{mdframed}[style=MyFrame]
\textbf{Summary to RQ2:} {\scshape SynCorr} can  more effectively repair regexes compared with the state-of-the-art tool {\scshape RFixer}. In addition, {\scshape SynCorr} and {\scshape RFixer} are complementary and simultaneously useful for our {\scshape TransRegex}.
\end{mdframed}
\begin{table}[thbp]
  \caption{The Number of Successful Repairs by {\scshape SynCorr} and {\scshape RFixer} on Three Datasets.}
  \label{tab:rq2}
  
  \renewcommand\arraystretch{1.2}
   \begin{adjustbox}{width=1.0\linewidth,center}
\begin{tabular}{|l|c|c|c|}
     \hline
    \multirow{2}{*}{\textbf{Approach}} &\multirow{2}{*}{\textbf{KB13}}&\multirow{2}{*}{\textbf{NL-RX-Turk}}&\textbf{Structured} \\
    &&&\textbf{Regex}\\
    \hline
    \hline
    {\small{\scshape RFixer}} &25/45 (55.6\%)&780/930 (83.9\%)&245/712 (34.4\%)\\
    \hline
    {\small  {\scshape SynCorr}}&30/45 (66.7\%)& 785/930 (84.4\%)&346/712 (48.6\%)\\
     \hline
    {\small{\scshape RFixer} + {\scshape SynCorr}}&35/45 (77.8\%)&895/930 (96.2\%)&387/712 (54.4\%)\\
    \hline
\end{tabular}
\end{adjustbox}
\end{table}

\begin{table*}[thbp]
\renewcommand\arraystretch{1.8}
\definecolor{CornflowerBlue}{HTML}{1F4E79}
  \definecolor{LimeGreen}{HTML}{548235}
  \definecolor{col1}{HTML}{B4C7E7}
  \definecolor{col2}{HTML}{8FAADC}
  \definecolor{cRed}{HTML}{C00001}%
  \definecolor{cGreen}{HTML}{00CD66}
  \definecolor{lightyellow}{rgb}{1.0, 1.0, 0.88}
\centering
  \caption{Examples \chm{of Repair} \iffalse Incorrect Regexes Repaired\fi by {\scshape RFixer} and {\scshape SynCorr}.}\label{tab:repair}
 \begin{adjustbox}{width=1.0\linewidth,center}
\begin{tabular}{|l|p{0.29\textwidth}|p{0.29\textwidth}|p{0.29\textwidth}|p{0.29\textwidth}|}
    \hline
    
     \textbf{No.} &\textbf{Incorrect Regex}&\textbf{Ground Truth}&\textbf{{\scshape RFixer}}&\textbf{{\scshape SynCorr}}\\
     \hline
   \hline
   \#1&([AEIOUaeiou].*[0-9].*)\textbf{\textcolor{cRed}{\underline{\{7,\}}}}&[AEIOUaeiou].*[0-9]\textbf{\textcolor{CornflowerBlue}{\underline{\{7,\}}}}.*&([AEeiO01234U56789].*[0-9].*)\{2,\}  \quad \textcolor{cRed}{\xmark}&[AEIOUaeiou].*[0-9]\{7,\}.*\quad \textcolor{LimeGreen}{\cmark}\\
   \hline
  \#2&[A-Za-z]\{2,3\}[a-z]\textbf{\textcolor{cRed}{\underline{\{2,3\}}}}[A-Z]\{3,4\}&
  [A-Za-z]\{2,3\}[a-z]\textcolor{CornflowerBlue}{\underline{\{3\}}}[A-Z]\{3,4\}&
  ([AabBCDEeFGHIJLnXxy]\{2,\}[abcdefl npvwxy])+[a-z]\{2,3\}[A-Z]\{3,4\} \quad \textcolor{cRed}{\xmark}&
  [A-Za-z]\{2,3\}[a-z]\textbf{\textcolor{LimeGreen}{\underline{\{3,3\}}}}[A-Z]\{3,4\} \quad \textcolor{LimeGreen}{\cmark}\\
   \hline
   \#3&\textbf{\textcolor{cRed}{\underline{([A-Z]$|$[a-z])\{1,\}}}}\&.\{6,8\}\&(.*([A-Z]$|$[a-z]).*)&.\{6,8\}\&(.*[A-Za-z].*)&\textbf{Not supporting regexes with \&}&.\{6,8\}\&(.*([A-Z]$|$[a-z]).*)\quad \textcolor{LimeGreen}{\cmark}\\
   \hline
   \#4&\textbf{\textcolor{cRed}{\underline{[A-Za-z]}}}\{3,\}[0-9]\{3,\}\textbf{\textcolor{cRed}{\underline{N}}}[A-Za-z]\{2,4\}&\textbf{\textcolor{LimeGreen}{\underline{[A-Z]}}}\{3,\}[0-9]\{3,\}\textbf{\textcolor{LimeGreen}{\underline{(N$|$g)}}}[A-Za-z]\{2,4\}&\textbf{\textcolor{LimeGreen}{\underline{[A-Z]}}}\{3,\}[0-9]\{3,\}\textbf{\textcolor{LimeGreen}{\underline{[gN]}}}[a-zA-Z])\{2,4\}\quad \textcolor{LimeGreen}{\cmark}&\textbf{Trapping in local optimum}\\
 \hline
\end{tabular}
\end{adjustbox}
\end{table*}


\begin{table*}[thbp]
\renewcommand\arraystretch{1.8}
\definecolor{CornflowerBlue}{HTML}{1F4E79}
  \definecolor{LimeGreen}{HTML}{548235}
  \definecolor{col1}{HTML}{B4C7E7}
  \definecolor{col2}{HTML}{8FAADC}
  \definecolor{cRed}{HTML}{C00001}%
  \definecolor{cGreen}{HTML}{00CD66}
  \definecolor{lightyellow}{rgb}{1.0, 1.0, 0.88}
\centering
  \caption{Examples \chm{Illustrating the Benefits of Leveraging both Natural Language and Examples} \iffalse Successful Cases Generated by\fi \chm{in} {\scshape TransRegex}.}\label{tab:succ}
\begin{adjustbox}{width=1.0\linewidth,center}
\begin{tabular}{|l|p{0.45\textwidth}|p{0.2\textwidth}|p{0.2\textwidth}|p{0.2\textwidth}|}
    \hline
    
     \textbf{No.} &\textbf{Description}&\textbf{Ground Truth}&\textbf{\lyt{{\scshape $\rm S_2RE$}}}&\textbf{Success Type}\\
     \hline
   \hline
   
  \#1&{\footnotesize items with a capital letter preceding “dog” \textbf{\textcolor{cRed}{\underline{at least 3 times}}}}&([A-Z].*dog.*)\textbf{\textcolor{CornflowerBlue}{\underline{\{3,\}}}}&[A-Z].*(dog)\textbf{\textcolor{LimeGreen}{\underline{\{3,\}}}}.*&Ambiguity of NL\\
  \hline
  \#2&{\footnotesize the string should start with at least 1 or more \textbf{\textcolor{cRed}{\underline{capital i}}}, then it is followed by 3 letters.} &\textbf{\textcolor{CornflowerBlue}{\underline{l\{1,\}}}}[A-Za-z]\{3\}&\textbf{\textcolor{LimeGreen}{\underline{[A-Z]\{1,\}}}}[A-Za-z]\{3\}& Imprecision of NL\\
  \hline
  \#3&{\footnotesize the string must begin with 2 or more letter s. after this grouping, it is a 3 digit number. after this 3 digit number, there a letter Z. after the letter Z, the string must contain \textcolor{cRed}{\underline{$\bm{\_;}$}} or \textcolor{cRed}{\underline{$\bm{*\#\#}$}}. the string can end with an optional string of 3 to 4 capital letters.} &{\footnotesize s\{2,\}[0-9]\{3\}Z\textbf{\textcolor{CornflowerBlue}{\underline{($\bm{\_;}$$\bm{|}$$\bm{*\#\#}$)}}}([A-Z]\{3,4\})?}&{\footnotesize s\{2,\}[0-9]\{3\}Z\textbf{\textcolor{LimeGreen}{\underline{[0-9]\{3\}}}}([A-Z]\{3,4\})?}&Unknown or rare words\\
\hline
    \#4 &{\footnotesize a list of 3 comma separated strings of lowercase letters.} &  [a-z]+\textcolor{CornflowerBlue}{\underline{(,[a-z]+)(,[a-z]+)}}& [a-z]+\textcolor{LimeGreen}{\underline{(,[a-z]+)*}}&False prediction by \lyt{{\scshape $\rm S_2RE$}}\\
 \hline
\end{tabular}
\end{adjustbox}
\end{table*}

\subsection{RQ3: Effectiveness  of TransRegex}\label{sec:rq3}
To answer this question, we compared {\scshape TransRegex} with six NLP-based  baselines and four multi-modal baselines. 
We can see from TABLE~\ref{tab:accuracy} that on average, NLP-based works performed worse than multi-modal works on all three datasets. In general, the accuracy on the first dataset KB13 is much higher than that on the other two datasets \chm{for NLP-based methods}, and the accuracy achieved by NLP-based methods are up to 30\% lower than that achieved by multi-modal methods \chm{on average}. Particularly, on the first two datasets, the accuracy of NLP-based works range from 38.6\% to 78.2\%, compared with 68.9\% to 86.4\% achieved by existing multi-modal works.
On comparison, {\scshape TransRegex} achieved approximate 10\% higher accuracy than these baselines, reaching \lyt{90.3\% to 98.6\%} \chm{on the first two datasets}.
The superiority of our work is more obvious on the last dataset, which is much more complex than the first two. 
The accuracy achieved by {\scshape TransRegex} (67.4\%) almost doubled the accuracy achieved by {\scshape Deep-Regex} (Ye et al.) + {\scshape Exs} (37.2\%). 




Let us present a few \chm{example regexes that are synthesized incorrectly from \lyt{NL} \iffalse natural language\fi descriptions but are synthesized correctly by our {\scshape TransRegex}} to illustrate the benefits of leveraging both \lyt{NL} and examples.
As case \#1 in TABLE~\ref{tab:succ}, the description is ambiguous, i.e.,  it is unclear what part of the string would appear at least 3 times, resulting in that \lyt{{\scshape $\rm S_2RE$}} predicted a regex embedding other meanings.
Our {\scshape TransRegex} utilizes examples (e.g., a positive example \verb|AdogBdogCdog|) to help
disambiguate the description, and fix the incorrect regex \verb|[A-Z].*(dog){3,}.*| to the correct regex \verb|([A-Z].*dog.*){3,}|.
Similar to cases \#2, \#3, and \#4 in TABLE~\ref{tab:succ}, the errors that are caused by imprecision/unknown words in descriptions or false prediction by NLP-based approaches can be corrected by {\scshape TransRegex} through using the example-guided regex repairer.

%

Further, we analyze the possible reasons why the other multi-modal works are not as effective as ours as follows.
As mentioned above, {\scshape Deep-Regex} (Ye et al.) + {\scshape Exs}
simply uses examples to select the $k$-best regexes among the candidates produced by {\scshape Deep-Regex} (Ye et al.).
If there is no correct regex in the candidates, the examples will not help.
The three variants algorithms (i.e., {\scshape GrammarSketch}+ MLE, {\scshape DeepSketch}+ MLE, and {\scshape DeepSketch}+ MML) rely heavily on the quality of the sketches synthesized in their first step. 
In other words, the incorrection of sketches will be inherited by the generated regex in the next step. While  {\scshape TransRegex} does not have the above-mentioned drawbacks. 
In addition, although  {\scshape TransRegex} is also a two-step algorithm, the second step of {\scshape TransRegex} is not only not affected by the errors of the first step, but also specifically correct the errors of the first step guided by the given examples.


\begin{mdframed}[style=MyFrame]
\textbf{Summary to RQ3:} {\scshape TransRegex} can achieve higher accuracy than the NLP-based works with  17.4\%, 35.8\% and 38.9\%, and the state-of-the-art multi-modal works with 10\% to 30\% higher accuracy on all three datasets. The experiment results also indicate {\scshape TransRegex} utilizing natural language and examples in a more effective way than other multi-modal works.
\end{mdframed}

\subsection{RQ4: Efficiency of TransRegex}\label{sec:rq4}

\begin{table}[thbp]
  \caption{Average running time per benchmark on three datasets.}
  \label{tab:efficiency}
 \renewcommand\arraystretch{1.2}
  \begin{adjustbox}{width=1.0\linewidth,center}
\begin{tabular}{|l|c|c|c|ccc}
\hline
 \multirow{2}{*}{\textbf{Approach}} &\multirow{2}{*}{\textbf{KB13}}&\multirow{2}{*}{\textbf{NL-RX-Turk}}&\textbf{Structured} \\
    &&&\textbf{Regex}\\
    \hline
    \hline
    {\small{\scshape Deep-Regex} (Locascio et al.)} &2.621 s& 1.104 s &2.108 s\\
    {\small\lyt{{\scshape $\rm S_2RE$}}}&3.578 s& 1.656 s &3.313 s\\
    \hline
    \hline


    {\small{\scshape TransRegex} (\lyt{{\scshape $\rm S_2RE$}} + {\scshape SynCorr})}&4.958 s&3.085 s&8.624 s\\
    {\small{\scshape TransRegex} (\lyt{{\scshape $\rm S_2RE$}} + {\scshape RFixer})}&5.821 s&4.737 s&22.460 s\\
     {\small{\scshape TransRegex} (\lyt{{\scshape $\rm S_2RE$}} + {\scshape SynCorr} + {\scshape RFixer})}&6.688 s&4.011 s&13.737 s\\
\hline
\end{tabular}
\end{adjustbox}
\end{table}

To evaluate the efficiency of three variants of {\scshape TransRegex}, we compared the average running time of synthesizing per regex with the open-source baseline {\scshape Deep-Regex} (Locascio et al.) and our \lyt{{\scshape $\rm S_2RE$}} in Table~\ref{tab:efficiency}. In general, we can see that \lyt{{\scshape TransRegex}  takes longer time to synthesize regexes on the last dataset than on the first two, and NLP-based baselines take less than {\scshape TransRegex} on average.}
In particular, {\scshape TransRegex} takes an average 5.822 seconds and 3.944 seconds on the first two datasets, compared with 1.104 to 3.578 seconds achieved by baselines. While on the last dataset, {\scshape TransRegex} takes longer time due to the complexity of the dataset. 
Considering together with the accuracy, {\scshape TransRegex} achieved the accuracy (67.4\%) that doubles the accuracy (28.5\%) achieved by \lyt{{\scshape $\rm S_2RE$}}, taking only around 10 more seconds running time. 
\chm{Table~\ref{tab:efficiency} also reveals the rationality of our algorithms,} the adopting of {\scshape SynCorr} helps us to accelerate the repair process in some cases, while {\scshape Rfixer} takes care of the rest.

\begin{mdframed}[style=MyFrame]
\textbf{Summary to RQ4:} {\scshape TransRegex} can synthesize regex efficiently. Especially when considering together with accuracy, {\scshape TransRegex} can takes an average 3.944 to 5.822 seconds to achieve around 20\% more accuracy on simpler datasets, and takes 30\% more accuracy at the cost of 10 more seconds on the more complex dataset.
\end{mdframed}

\begin{table*}[thbp]
\renewcommand\arraystretch{1.3}
\definecolor{CornflowerBlue}{HTML}{1F4E79}
  \definecolor{LimeGreen}{HTML}{548235}
  \definecolor{col1}{HTML}{B4C7E7}
  \definecolor{col2}{HTML}{8FAADC}
  \definecolor{cRed}{HTML}{C00001}%
  \definecolor{cGreen}{HTML}{00CD66}
  \definecolor{lightyellow}{rgb}{1.0, 1.0, 0.88}
\centering
  \caption{Examples \chm{of} failed cases generated by {\scshape TransRegex}.}\label{tab:fail}
  \begin{adjustbox}{width=1.0\linewidth,center}
\begin{tabular}{|l|p{0.24\textwidth}|p{0.188\textwidth}|p{0.18\textwidth}|p{0.22\textwidth}|p{0.07\textwidth}|p{0.22\textwidth}|p{0.15\textwidth}|}
    \hline
    
     \textbf{No.} &\textbf{Description}&\textbf{Positive Examples}&\textbf{Negative Examples}&\textbf{Ground Truth}&\multicolumn{2}{|c|}{\textbf{Predicted Results}}&\textbf{Failure Type}\\
     \hline
   \hline
  
  &&\verb|fsuhoRdKRUGrFIRj|&\verb|qmnVF|&&{\scshape $\rm S_2RE$} &\verb/[a-z]{3}[A-Za-z]{3,}/&\\
  &&&&& &&\\
  \cline{6-7}
  \multirow{2}{*}{\#1}&{\footnotesize  3 lower case letters followed by more }&\verb|iptHLAdpPnKUXPrWo|&\verb|qmnVa|&\verb/[a-z]{3}[A-Za-z]{3,}/&  {\scshape SynCorr} &\verb/[a-z]{3}[A-Za-z]{3,}[A-Z/&\textbf{Uncharacteristic}\\
   &{\footnotesize  than 3 letters.}&&&&&\verb/a-z]{3,}/&\textbf{examples}\\
\cline{6-7}
  &&$\ldots$&$\ldots$&&{\scshape RFixer} &\verb/[a-z]{3}[A-Za-z]{3,}/&\\
   &&&&& &&\\
  \hline

  &{\footnotesize a string that consists of upper or lower }&\verb|==;0#=#+k0-0| \quad \textbf{\textcolor{cRed}{\xmark}}&\verb|!;_|&&\lyt{{\scshape $\rm S_2RE$}}&\verb/([A-Za-z]|[-!@#$%^&*()_.]/&\\
   &{\footnotesize case letters, special characters} \verb/(-!@#/&&&&&\verb/|0){1,}&.{4,}/&\\
   \cline{6-7}
   
  \multirow{2}{*}{\#2}&\verb|$%^&*()_.)| {\footnotesize or the number 0 and } &\verb|sy=;-0M0=!0T0f!;E| \quad \textbf{\textcolor{cRed}{\xmark}} &\verb|#&0-!7!Kx;|&\verb/([A-Za-z]|[-!@#$%^&*()_.]/&{\scshape SynCorr}&\verb/.{4,}&(~([-!@#$%^&*()_.]/&\multirow{2}{*}{\textbf{Wrong examples}}\\
  
   &{\footnotesize whose length is 4 or more characters} &&&\verb/|0){1,}&.{4,}/&&\verb/.*))/&\\
    \cline{6-7}
    
  &{\footnotesize  long.}&$\ldots$&$\ldots$& &{\scshape RFixer}&Not supporting regexes with \&&\\
   & &&& &&&\\
  \hline
  &{\footnotesize a list of 3 semicolon separated strings,}&\verb|O0Culdvs;0ctycf;H|&\verb|mLNWmxydw;x9e5pdvf;A|&&\lyt{{\scshape $\rm S_2RE$}}&\verb/([A-Za-z]|[0-9]{1,})&.{4}/&\\
  &{\footnotesize  the first and second strings begin with }&&&\verb/([A-Za-z]|[0-9]){4}[a-z]/&&\verb/;[a-z]{2,4};[A-Z]{1,}/&\\
  \cline{6-7}
  
   \multirow{2}{*}{\#3}&{\footnotesize  any combination of 4 letters or digits,} &\verb|80Oymq;2T2myvsz;DZG|&\verb|5xSxe;xTC8jv;Q|&\verb/{2,4};([A-Za-z]|[0-9]){4}/&{\scshape SynCorr}&Time out!&\textbf{\lyt{Very} complex regexes}\\
&{\footnotesize  these strings end with 2 to 4 lower } &&&\verb/[a-z]{2,4};[A-Z]{1,}/&&&\textbf{or descriptions}\\
 \cline{6-7}

&{\footnotesize case letters, the third part is any number of capital letters.} &$\ldots$&$\ldots$&&{\scshape RFixer}&Not supporting regexes with \&&\\

 \hline
\end{tabular}
\end{adjustbox}
\end{table*}


\section{Threats to TransRegex's Validity}

{\scshape TransRegex} is not guaranteed to generate correct regexes for each benchmark, mainly due to the following aspects: 

\begin{itemize}
    \item \textbf{Uncharacteristic examples.} Although {\scshape TransRegex} has greatly reduced the amount of  required examples, the quality of {\scshape TransRegex} still depends on characteristic examples. When the examples provided by users are not characteristic, it is difficult for  {\scshape TransRegex} to get the correct regexes. For instance, the positive examples from case \#1 in TABLE~\ref{tab:fail} belongs to both the incorrect one   \verb|[a-z]{3}[A-Za-z]{3,}[A-Za-z]{3,}| and  ground truth \verb|[a-z]{3}[A-Za-z]{3,}|, i.e., fail to distinguish between the incorrect one and  ground truth. \chm{This may cause} {\scshape SynCorr} or {\scshape RFixer} to fail to repair the incorrect regex. If the user further provides examples (e.g.,  
   a positive example \verb|aaaAAA|) that can distinguish the two ones, our methods can easily 
   get the correct one.
    \item \textbf{Wrong examples.} Wrong examples provided by users, resulting in that our algorithms are misled to synthesize regexes in the wrong direction.
    More concretely, our algorithms repair incorrect regexes in the wrong direction and even fix the correct regexes produced by \lyt{{\scshape $\rm S_2RE$}} into the incorrect one. As case \#2 in TABLE~\ref{tab:fail}, if we only use the \lyt{NL} description, we can get the accurate regex \verb/([A-Za-z]|[-!@#$%^&*()_.]|0){1,}&.{4,}/, which is the same as the ground truth.
    However, the user provides some wrong examples (e.g., a positive example \verb|==;0#=#+k0-0|), which leads our algorithms to believe that the regex \verb/([A-Za-z]|[-!@#$%^&*()_.]|0){1,}&.{4,}/ is not accurate, and then our algorithm uses the wrong examples to fix the accurate regex into an incorrect one.
    \item \textbf{\chm{Very} complex regexes or descriptions.} Similar to case \#3 in TABLE~\ref{tab:fail}, if the \lyt{NL} description is very long and complicated, or 
    the target regex is very complex in terms of length and tree-depth, the regex synthesized by \lyt{{\scshape $\rm S_2RE$}} may be very different from the target one. This kind of regex, which is very different from the target one and required very large fixes, is difficult for our algorithms to repair into a correct one in a limited time and space.
\end{itemize}

\section{Related Work}\label{related_work}
\subsection{Regex Synthesis}

\noindent\textbf{Regex synthesis from examples.}
The problem of automatic regex synthesis from examples has been explored in many domains~\cite{DBLP:conf/gpce/LeeSO16,DBLP:journals/computer/BartoliDLMS14,DBLP:journals/tkde/BartoliLMT16,DBLP:journals/tods/BexNSV10,DBLP:journals/tweb/BexGNV10,DBLP:conf/dasfaa/LiZCCG19,DBLP:journals/mst/FreydenbergerK15,ase20}. 
AlphaRegex~\cite{DBLP:conf/gpce/LeeSO16} is a search-based algorithm for synthesizing simple regexes for introductory automata assignments.
AlphaRegex exploits over/under-approximations to effectively prune out a large search space.
However, all the regexes produced by AlphaRegex are over alphabets of size 2. 
{\it RegexGenerator++}~\cite{DBLP:journals/computer/BartoliDLMS14,DBLP:journals/tkde/BartoliLMT16} is a state-of-the-art approach for the synthesis of regexes from positive and negative examples. 
The fact that {\it RegexGenerator++} utilizes genetic programming means that it is not guaranteed to generate a correct solution--\emph{i.e.}, accepting all the positive examples while rejecting all the negative examples.
Lots of existing works focus on XML schemas inference~\cite{DBLP:journals/tods/BexNSV10,DBLP:journals/tweb/BexGNV10,DBLP:conf/dasfaa/LiZCCG19,DBLP:journals/mst/FreydenbergerK15}, via resorting to infer regexes from examples. These approaches usually aim to tackle restricted forms of 
regexes from positive examples only.
\lyt{Li et al.~\cite{ase20} presented a novel algorithm {\scshape FlashRegex} to generate anti-ReDoS regexes from given positive and negative examples by reducing the ambiguity of these regexes and using SAT techniques.}

However, one of the main issues in the above example-based techniques is the quality of the synthesis, i.e., whether it would generalize and correct for unseen examples.
Specifically, if users can not provide sufficient and characteristic examples, the synthesized regexes will be under-fitting or over-fitting.

\noindent\textbf{Regex synthesis from \lyt{NL}.}
Several works from the Natural Language Processing (NLP) community address the problem of generating regexes from \lyt{(NL)} specifications~\cite{DBLP:conf/naacl/KushmanB13,DBLP:conf/emnlp/LocascioNDKB16,DBLP:conf/emnlp/ZhongGYPXLLZ18,park-etal-2019-softregex}.
Kushman and Barzilay~\cite{DBLP:conf/naacl/KushmanB13} introduced a technique for learning
a probabilistic combinatory categorial grammar  model to
parse a \lyt{NL} description into a regex.
To avoid domain-specific feature extraction, Locascio et al.~\cite{DBLP:conf/emnlp/LocascioNDKB16} described the {\scshape Deep-Regex} model based on standard sequence-to-sequence (seq2seq) model, which \lyt{regards} the problem of generating regexes from \lyt{NL} descriptions as a direct {\it machine translation} task.
To solve the problem that {\scshape Deep-Regex} model may not generate semantically correct regexes, the {\scshape SemRegex} model~\cite{DBLP:conf/emnlp/ZhongGYPXLLZ18} based on reinforcement learning method was presented. \chm{It} leverages DFA equivalence as a reward function to encourage the model to generate semantically correct regexes. To speeds up the training phase of {\scshape SemRegex} model, Park et al.~\cite{park-etal-2019-softregex} devised the {\scshape SoftRegex} model, which determines the equivalence of two regexes using deep neural networks. 

There are three major bottlenecks in existing NLP-based techniques that affect the quality of synthesis:
(\rmnum{1}) Ambiguity and imprecision of \lyt{NL} . Ambiguity of \lyt{NL} results in predicting a regex embedding other meanings (resp. imprecision of \lyt{NL} affects the correctness of synthesis); 
(\rmnum{2}) Unknown words and rare words.  
There are unknown words or rare words in \lyt{NL} descriptions, which will lead to failure to generate correct regexes;
(\rmnum{3}) Seq2Seq model. Seq2Seq-based approaches can only synthesize regexes similar in shape to the training data.

\noindent\textbf{Regex synthesis from \lyt{NL} \iffalse natural language\fi and examples.}
Ye et al.~\cite{DBLP:conf/acl/YeCDD20} proposed StructuredRegex, a new dataset for regex synthesis from \lyt{NL} 
and examples.
Ye et al.~\cite{tacl} introduced a baseline model {\scshape Deep-Regex} + {\scshape Filter}, which uses {\scshape Deep-Regex} as base model, and considers examples by simply filtering the $k$-best regexes. However, the positive and negative examples are not considered in the training and inference phase.
The latest two works~\cite{tacl,DBLP:conf/acl/YeCDD20} presented new two-step frameworks for regex synthesis from \lyt{NL} and examples. First, a semantic parser  converts the \lyt{NL} description
into an intermediate sketch. Then a synthesizer searches the regex space defined by the sketch and returns a concrete regex that is consistent with the given examples. Although both adopting a two-step paradigm, these works~\cite{tacl,DBLP:conf/acl/YeCDD20} have an apparent limitation---incorrect sketches generated in \chm{the} first step will subsequently induce the final \chm{regexes}. In other \chm{words}, the incorrection of sketches will be inherited by the synthesized \chm{regexes} in the next step. On the other hand, our work overcomes this limitation: the second step of {\scshape TransRegex} (i.e., \emph{example-guided regex repair}) fixes the incorrect regex by examples if needed. \chm{In this manner}, the inherited inconsistencies in our first step (i.e., \emph{NLP-based regex synthesis}) will be fixed in our second step. 

\subsection{Regex Repair}
\noindent\textbf{Regex repair from examples.}
There are several works~\cite{DBLP:conf/emnlp/LiKRVJ08,DBLP:conf/pakdd/RebeleTS18,DBLP:journals/pacmpl/PanHXD19,ase20} targeting at repairing regexes from examples. We discuss two main paradigms of them. In the first paradigm, works only consider either positive or negative examples. Li et al.~\cite{DBLP:conf/emnlp/LiKRVJ08} proposed ReLIE, which can modify complex regexes by rejecting the newly-input negative examples. By contrast, Rebele et al.~\cite{DBLP:conf/pakdd/RebeleTS18} proposed a novel way to generalize a given regex so that it accepts the given positive examples. 
On the other hand, works in the second paradigm take both positive and negative examples into consideration. Pan et al.~\cite{DBLP:journals/pacmpl/PanHXD19} designed {\scshape RFixer}, a tool for repairing incorrect regexes using both examples. It took advantage of skeletons of regexes (i.e., sketches) to effectively prune out the search space, and it employed SMT solvers to efficiently explore the sets of possible character classes and numerical quantifiers. Our work applies \chm{{\scshape RFixer}} in our regex synthesis from \lyt{NL} and examples.
Li et al.~\cite{ase20} described algorithm {\scshape RepairingRE} based on Neighborhood
Search (NS) to repair incorrect or ReDos-vulnerable regexes from positive and negative examples. Similar to {\scshape RepairingRE}, our algorithm {\scshape SynCorr} also uses NS to repair regexes, but the difference is that {\scshape RepairingRE} uses automaton-directed repair, while we use regex-directed repair.

Like the above example-based synthesis algorithms, these repair algorithms also may cause under-fitting or over-fitting results.
To alleviate the problems of under-fitting/over-fitting, our {\scshape SynCorr} leverages some rewriting rules for sub-regexes abstraction to preserve the integrity of some small sub-regexes.

\subsection{Program Synthesis}
\noindent\textbf{Programming by example (PBE).} PBE techniques have been the subject of research in the past few decades~\cite{DBLP:conf/ijcai/ShawWG75} and successful paradigms for program synthesis, allowing end-users to construct and run new programs by providing examples of the intended program behavior~\cite{DBLP:conf/ijcai/EllisG17}.
Recently, PBE techniques have been successfully used for string transformations~\cite{DBLP:conf/popl/Gulwani11,DBLP:journals/pvldb/Singh16,DBLP:journals/pvldb/SinghG12}, data filtering~\cite{DBLP:conf/oopsla/WangGS16}, data structure manipulations~\cite{DBLP:conf/pldi/YaghmazadehKDC16,DBLP:conf/pldi/FeserCD15}, table transformations~\cite{DBLP:conf/pldi/FengMGDC17,DBLP:conf/pldi/HarrisG11}, SQL queries~\cite{DBLP:conf/pldi/WangCB17,DBLP:conf/kbse/ZhangS13},  and MapReduce programs~\cite{DBLP:conf/pldi/SmithA16,DBLP:journals/corr/AhmadC16}.

\noindent\textbf{Programming by NL (PBNL).}
There has been a lot of progress made in PBNL~\cite{DBLP:journals/ftpl/GulwaniPS17}. Specifically, several techniques have been proposed to translate \lyt{NL} descriptions into Python~\cite{DBLP:conf/acl/YinN17,DBLP:conf/acl/RabinovichSK17}, SQL queries~\cite{DBLP:conf/naacl/HuangWSYH18,DBLP:journals/pacmpl/Yaghmazadeh0DD17,DBLP:conf/ijcai/Zeng0GCLLTZ20}, shell scripts~\cite{DBLP:journals/ijseke/LiWYT19,DBLP:conf/lrec/LinWZE18}, spreadsheet formulas~\cite{DBLP:conf/sigmod/GulwaniM14}, test oracles~\cite{DBLP:conf/icse/MotwaniB19}, JavaScript function types~\cite{DBLP:conf/icse/MalikPP19}, and Java expressions~\cite{DBLP:conf/oopsla/GveroK15}.


\noindent\textbf{Program synthesis from \lyt{NL} \iffalse natural language\fi and examples.}
Sinece program synthesis from \lyt{NL}  and examples techniques can well overcome the shortcomings of PBE and PBNL techniques, at the same time, provide a more natural and friendly interface to the users, recent years they  have been widely used in several areas, for example, string manipulation programs~\cite{DBLP:conf/aaai/ManshadiGA13,DBLP:conf/ijcai/RazaGM15}, and  program sketches~\cite{DBLP:conf/icml/NyeHTS19}.
In this paper, we focus on an important subtask of the program synthesis from \lyt{NL} and examples problem: synthesizing regexes from both NL and examples.
\section{Conclusion}
\chm{We propose an automatic framework {\scshape TransRegex}, for synthesizing regular expressions from both natural language descriptions and examples. To the best of our knowledge, {\scshape TransRegex} is the first to treat the NLP-and-example-based regex synthesis problem as the problem of NLP-based synthesis with regex repair. For NLP-based synthesis, we devise a two-phase algorithm \lyt{{\scshape $\rm S_2RE$}} which generates more valid regexes while having similar or higher accuracy than the state-of-the-art NLP-based models. While for regex repair, we present a novel algorithm {\scshape SynCorr} that leverages NS algorithms to guide the search for a target regex and uses rewriting rules to \chm{alleviate} under-fitting/over-fitting and efficiently reduce the search space. 
The evaluation results demonstrate that the accuracy of our {\scshape TransRegex}  is 17.4\%, 35.8\% and 38.9\% higher than that of NLP-based works on the three publicly available datasets, respectively. Further, {\scshape TransRegex} can achieve higher accuracy than the state-of-the-art multi-modal works with 10\% to 30\% higher accuracy on all three datasets. The evaluation results also indicate {\scshape TransRegex} utilizing natural language and examples in a more effective way.}

\section*{Acknowledgment}

\lyt{The authors would like to thank the anonymous reviewers for their
comments and suggestions.
This work is supported in part by National Natural Science Foundation of China (Grants \#61872339, \#61472405, \#61932021, \#61972260, \#61772347, \#61836005), National Key Research and Development Program of China under Grant \#2019YFE0198100, Guangdong Basic and Applied Basic Research Foundation under Grant \#2019A1515011577, and Huawei PhD Fellowship, MSRA Collaborative Research Grant.
}

\newpage
\bibliographystyle{IEEEtran}
\bibliography{icse21-full}

\end{document}